\documentclass[twocolumn,showpacs,showkeywords,preprintnumbers,amsmath,amssymb]{revtex4}
\usepackage{graphicx}
\usepackage{dcolumn}
\usepackage{epsfig,graphicx,amsmath}
\usepackage{hyperref}
\begin{document}
\title{Two-time second-order correlation function}

\author{Sintayehu Tesfa} \email{sint_tesfa@yahoo.com}
\affiliation{Physics Department, Addis Ababa University, P. O. Box 1176, Addis Ababa, Ethiopia}

\date{\today}

\begin{abstract}Derivation of two-time second-order correlation function by following approaches such as stochastic differential equation, coherent-state propagator, and quasi-statistical distribution function is presented. In the process, the time dependence of the operators is transferred to the density operator by making use of trace operation in which the coherent state propagator and $Q$-function that represent the quantum system under consideration are expressed in terms of different time parameters. Even though the number of resulting integrations are found to be large, the accompanying implementation turns out to be straightforward in view that the associated $c$-number functions are Gaussian by nature. In relation to the asserted possibility of rewriting the result of one of the approaches in terms of the other,  the presented derivation is expected to lay a strong foundation for viable technique of calculating correlations of various moments at different times that can be deployed in revealing quantum correlations. \end{abstract}

 \maketitle

\section{INTRODUCTION}

Photon measurement strategy designed for counting the number of photons lacks a potential for revealing correlation among photons generated at different times or positions; and so, would not be deployed for witnessing nonclassical traits of radiation \cite{n1781046}. Nonetheless, in case a light generated by a single source is orchestrated to travel over unequal distances, interference pattern can be casted at the site of the detector \cite{ajs34333}. Photons separated in time or position can also be counted by more than one detector--which makes exploring the nature and mechanism of evaluating correlation at different times or positions an integral part of optical experiments \cite{pr28491, jo19094002}. It may then suffice to mention Brown and Twiss \cite{n17727}, Slusher {\it{ et al.}} \cite{prl552409}, and Aspect {\it{ et al.}} \cite{prl491804} experiments in which coincidence and photon difference measurements unravel inherent nonclassical features such as squeezing, entanglement, and non-locality \cite{books}. Two-time second-order correlation function in particular is one of the parameters applicable to identify sources of light with non-classical properties; the attribute that makes it the most sought for in the field such as quantum optical processing  \cite{books, jmo541759} and spectroscopy \cite{oe244278}. In the same perspective, sending coherent light via or across different pathes or media using Match-Zehender interferometer so that coherent superposition can be induced also leads to quantum correlation as in early and late or Schr\"{o}dinger cat or NOON state \cite{prl852733} that would be the basis for quantum optical technology and enhanced measurement \cite{books}. 

This discussion warrants the importance of two-time second-order correlation to carry out and interpret results in research related to quantum optical setups \cite{pla435128059}. It is hence practically and pedagogically attractive looking for approaches that can be useful in deriving two-time second-order correlation  \cite{apro, pre562747}. Two-photon quantum correlation between radiations generated by lambda three-level atomic system for instance is studied applying the master equation formulation and Onsager-Lax regression theorem \cite{pra72043811}. Theoretical analysis of delayed coincidences of the radiation generated by two-level atom  placed in a cavity has also been addressed in view of different approaches \cite{jmo56105, pra414083}. Such works show that nonclassical attributes that include photon anti-bunching and sub-Poisson photon statistics can be observed. There has also been a great deal of interest in studying statistical and quantum traits of radiation generated by different schemes in terms of multi-time correlations for open and closed systems \cite{jpsj692873}. Since the solution of the density matrix  may not be sufficient to obtain two-time correlation function as usually presumed, it turns out to be compelling to resort to the transition probability distribution or employing the explicit form of one-time correlation function that can be calculated by applying master or Langevin or stochastic differential equation.  

To extend this consideration to more interesting cases, it is thought to be indispensable categorizing alternative approaches for calculating correlations among photons arriving at the site of detectors \cite{apro}. The main aim of this communication is hence directed towards filling the gap in this direction. In the first place, the general understanding of quantum correlation in relation to the seminal work of Glauber \cite{pr1302529} is presented. The technique of evaluating two-time correlation function by applying Lax-Onsager theorem \cite{lou, pr1292342, pr37405}, coherent-state propagator \cite{pra26451, jpamg347227}, and quasi-statistical distribution functions \cite{pr40749, zp461, pr1312766} is then explored, where the former is included for the sake of completeness. The latter however is anticipated to be of great utility in calculating various correlations that can be described in the form of moments of the radiation at different times. It so argued that to design, carry out, and analyze the results of implementations associated with correlation among various parameters, it is required to have at least the basic understanding of how the interrelation can be calculated, which makes this contribution valuable source to begin with.

\section{General Background}

It is common knowledge that theory of photon measurement requires proper characterization of the interaction of radiation with matter, which is not often considered due to the accompanying complications resulting from lack of complete knowledge of interaction \cite{books}. In this regard, since the  detectors are presumed to be insensitive to spontaneous emission, annihilation operator of the electric field is designated to represent counting process. In case the field undergoes transition from initial to final state, the elements of the transition matrix can be written  in the form
\begin{align}\label{5001}P_{T} = \langle f|\hat{E}^{(\dagger)}|i\big\rangle.\end{align}
On the other hand, when the measuring device is assumed to be ideal photon detector with frequency independent absorption probability, the total count or average field intensity is accounted for by summing overall states that can be occupied via absorption process \cite{pr1302529}:
\begin{align}\label{5006}I({\bf{r}},t) = \big\langle i|\hat{E}^{(-)}({\bf{r}},t)\hat{E}^{(\dagger)}({\bf{r}},t)| i\big\rangle.\end{align} 
It might be worthy noting that creation operator precedes the destruction operator--which indicates the notion of normal ordering.

This consideration insinuates that recording photon intensities using single detector may not ensure exhaustively measuring all properties of the field \cite{prl591903}. With this in mind, in case there is a field originating from position ${\bf{r}}$ and detected at separate times $t_{1}$ and $t_{2}$, the emerging correlation can be quantified by 
\begin{align}\label{5007}G^{(1)}({\bf{r}};t_{1},t_{2}) = {\rm{Tr}}\big(\hat{\rho} \hat{E}^{(-)}({{\bf{r}}},t_{1})\hat{E}^{(\dagger)}({\bf{r}},t_{2})\big),\end{align} 
where $\hat{\rho}$ is the density operator for radiation field \cite{apro}. It is not then hard to notice that Eq. \eqref{5007} stands for two-time first-order correlation function and found to be sufficient to embody classical like interference experiments that can be interpreted as the transition probability for the detector atom while it absorbs photon from a field at position ${\bf{r}}$ in time between $t$ and $t + dt$. 

It is a well established fact that stationary fields are common interest in quantum optics, that is, correlation function of the field is invariant under displacement of time variable. The correlation function $G^{(1)}({\bf{r}};t_{1},t_{2})$ is so taken to depend on $t_{1}$ and $t_{2}$ via their difference: $\tau = t_{2} - t_{1}$. The two-time first-order correlation function can then be expressed as $G^{(1)}({\bf{r}};\tau)$. In the same manner, the joint probability for detecting one photon at position ${\bf{r_{1}}}$ between $t_{1}$ and $t_{1} + dt_{1}$ and another at ${\bf{r_{2}}}$ between $t_{2}$ and $t_{2} + dt_{2}$ with $t_{1} < t_{2}$ is epitomized by two-time second-order correlation function \cite{pr1302529, loud}:
\begin{align}\label{5010}G^{(2)}({\bf{r}}_{1},{\bf{r}}_{2};t_{1},t_{2})&
 = {\rm{Tr}}\big(\hat{\rho}\hat{E}^{(-)}({\bf{r}}_{1},t_{1})\hat{E}^{(-)}({\bf{r}}_{2},t_{2})\notag\\&\times
\hat{E}^{(\dagger)}({\bf{r}}_{2},t_{2})\hat{E}^{(\dagger)}({\bf{r}}_{1},t_{1})\big),\end{align}
which can be perceived as delayed coincidences between two sets of radiation whose right hand  is time ordered: operators at earlier times come first, and are also normally ordered since creation operator comes first \cite{jac50357}. 

Normalized correlation function expressed in terms of radiation or boson operators is often demanded in quantum optics. In this consideration, the first-order normalized correlation function turns out in view of the relation between field and boson operators to be \cite{books}
\begin{align}\label{5011}g^{(1)}(\tau) =\frac{\langle\hat{a}^{\dagger}(t)\hat{a}(t+\tau)\rangle}{
\langle\hat{a}^{\dagger}(t)\hat{a}(t)\rangle}.\end{align} 
The second-order normalized two-time correlation function can also be expressed as
\begin{align}\label{5012}g^{(2)}(\tau) =\frac{\langle\hat{a}^{\dagger}(t) \hat{a}^{\dagger}(t+\tau) \hat{a}(t+\tau) \hat{a}(t)\rangle}{\langle\hat{a}^{\dagger}(t)\hat{a}(t)\rangle^{2}}\end{align}
and reduces at equal time to
\begin{align}\label{5012}g^{(2)}(0)={\langle\hat{a}^{\dagger}(t)\hat{a}(t)\rangle\big[\langle\hat{a}^{\dagger}(t) \hat{a}(t)\rangle-1\big]\over \langle\hat{a}^{\dagger}(t)\hat{a}(t)\rangle^{2}}.\end{align}

It might be essential emphasising that normalized correlation function is one of the tools applicable in identifying the inherent photon statistics \cite{oe244278, books, pla435128059}. Particularly when the light under study satisfies the inequality,
\begin{align}\label{5013}g^{(2)}(\tau) < g^{(2)}(0),\end{align}
it is deemed as exhibiting excess correlation. This characteristic denotes the phenomenon of photon bunching as the photons tend to distribute themselves in bunches rather than at random. When such radiation field falls on the detector more pairs of photons are detected closer together than further apart. Contrary to this, the phenomenon of photon anti-bunching is one of the possible mechanisms by which the accompanying nonclassical features are demonstrated as when two-level atom placed in a cavity interacts with radiation \cite{jmo541759, jmo56105}. It is also common practice categorizing photon statistics via calculating two-time second-order correlation function in which  $g^{(2)}(\tau) = 1$ represents Poissonian,  $g^{(2)}(\tau) > 1$ super-Poissonian, and $g^{(2)}(\tau) < 1$ sub-Poissonian  photon statistics \cite{books, pla435128059}.

Correlation function can also be explored by designating the time evolution of the accompanying operator as a
solution of Heisenberg's or quantum Langevin's equation--which might not always be the best choice. To obtain the two-time expectation value, it is so desirable to look for the utility that makes easier to calculate two-time higher-order correlations \cite{rmp3225, pr1292342, pr37405}. To begin with, one may argue that to get noise spectrum and photon count distribution,  the one-time solution of the dynamical equation provides the time-dependent information to determine the higher-order correlations \cite{oc179463}. In this regard, the density operator at a time $\tau$ with $\tau \geq 0$ is expressed in terms of density operator at earlier time (specially at $t = 0$) as
\begin{align}\label{5039}\hat{\rho}(\tau) = \hat{U}(\tau)\hat{\rho}(0)\hat{U}^{\dagger}(\tau).\end{align}
It may not be hard to see that $\hat{U}(\tau)$ is the evolution operator defined by
\begin{align}\label{5040}\hat{U}(\tau) = \exp(-i\tau\hat{H}_{S}),\end{align}
where $\hat{H}_{S}$ designates the system.

In Makrovian approximation, in which the interconnectedness between the states of the system and reservoir at equal time is presumed to be unimportant, the evolution of a single-time expectation value can be expressed in view of cyclic property of trace operation as
\begin{align}\label{tt1}\big\langle\hat{A}(t+\tau)\big\rangle&= {\rm{Tr}}_{S}\big[\hat{A}(t)\notag\\&\times {\rm{Tr}}_{R}\big(\hat{U}(\tau)\hat{\rho}_{S}(t)\otimes\hat{\rho}_{R}(t)\hat{U}^{\dagger}(\tau)\big)\big].\end{align}
Since $\hat{\rho}_{S}(t)\otimes\hat{\rho}_{R}(t)$ denotes density operator that represents combined system, it is possible to write
\begin{align}\label{tt2}\big\langle\hat{A}(t+\tau)\big\rangle&= {\rm{Tr}}_{S}\big[\hat{A}(t)\notag\\&\times {\rm{Tr}}_{R}(\hat{\rho}_{S}(t+\tau)\otimes\hat{\rho}_{R}(t+\tau)\big)\big],\end{align} 
which can also be put in the form
\begin{align}\label{5041}\big\langle\hat{A}(t+\tau)\big\rangle = \sum_{j}G_{j}(\tau)\langle\hat{A}_{j}(t)\rangle,\end{align}
where
\begin{align}G(\tau) = {\rm{Tr}}_{R}\big(\hat{\rho}_{S}(t+\tau)\otimes\hat{\rho}_{R}(t+\tau)\big)\end{align}
is the coefficient that accounts for the environment \cite{pr1292342, pr37405}. 

In the same manner, the two-time correlation function can be expressed in view of cyclic property of trace operation as
\begin{align}\label{tt4}\big\langle\hat{A}(t+\tau)\hat{B}(t)\big\rangle& ={\rm{Tr}}_{S}\big[\hat{A}(t)\hat{B}(t)\notag\\&\times{\rm{Tr}}_{R}\big(\hat{U}(\tau)\hat{\rho}_{S}(t)
\otimes\hat{\rho}_{R}(t)\hat{U}^{\dagger}(\tau)\big)\big].\end{align}
Comparison with earlier discussion then implies that
\begin{align}\label{5042}\big\langle\hat{A}(t+\tau)\hat{B}(t)\big\rangle = \sum_{j}G_{j}(\tau)\big\langle\hat{A}_{j}(t)\hat{B}_{j}(t)\big\rangle.\end{align}
What then remains to obtain two-time second-order correlation is finding the corresponding single-time correlation between variables of the system and the coefficient that epitomizes the environment, where the latter is more difficult to realize and controversial to interpret \cite{prl77798}.

\section{Some illustrative examples}

Time dependent density matrix may not be always sufficient to obtain two-time second-order correlation function; and so, it might be required to resort to the transition probability distribution or employing the explicit form of one-time correlation function that can be calculated employing master or Langevin or stochastic differential equation and so on. In addition to these intuitive approaches, more rigorous mathematical approaches are often demanded to determine correlation between beams of light arriving at measuring site specially when the system is coupled to environment \cite{pra97052101}

\subsection{Stochastic differential equation}

To demonstrate one of the approaches, one can begin with driving stochastic differential equation associated with the normal ordering for lasing mechanism in which degenerate three-level atoms in cascade configuration and initially prepared in coherent superposition of the upper and lower energy levels are injected at constant rate into the cavity \cite{arx, books}.  For this system, it is possible to see that 
\begin{align}\label{dt17}\frac{d}{dt}\langle\hat{a}(t)\rangle = -\frac{\mu}{2}\langle\hat{a}(t)\rangle
+ \beta\langle\hat{a}^{\dagger}(t)\rangle,\end{align}
\begin{align}\label{dt18}\frac{d}{dt}\big\langle\hat{a}^{2}(t)\big\rangle = -\mu\big\langle\hat{a}^{2}(t)\big\rangle +
2\beta\big\langle\hat{a}^{\dagger}(t)\hat{a}(t)\big\rangle-B,\end{align}
\begin{align}\label{dt19}\frac{d}{dt}\big\langle\hat{a}^{\dagger}(t)\hat{a}(t)\big\rangle &=
-\mu\big\langle\hat{a}^{\dagger}(t)\hat{a}(t)\big\rangle \notag\\&+
\beta\big[\big\langle\hat{a}^{\dagger^{2}}(t)\big\rangle +\big\langle\hat{a}^{2}(t)\big\rangle\big] + C,\end{align} 
where $\mu$, $\beta$, $B$, and $C$ are constants that account for the system and environment \cite{arx, pra74043816}. 

It is possible to observe that operators in Eqs. \eqref{dt17}, \eqref{dt18}, and \eqref{dt19} are in the normal order. Since the expectation values are essentially $c$-number, these expressions can be rewritten in terms of $c$-number variables associated with the normal ordering as
\begin{align}\label{dt22}\frac{d}{dt}\langle\alpha(t)\rangle = -\frac{\mu}{2}\langle\alpha(t)\rangle
+ \beta\langle\alpha^{*}(t)\rangle,\end{align}
\begin{align}\label{dt23}\frac{d}{dt}\big\langle\alpha^{2}(t)\big\rangle = -\mu\big\langle\alpha^{2}(t)\big\rangle +2\beta\big\langle\alpha^{*}(t)\alpha(t)\big\rangle-B,\end{align}
\begin{align}\label{dt24}\frac{d}{dt}\big\langle\alpha^{*}(t)\alpha(t)\big\rangle &=
-\mu\big\langle\alpha^{*}(t)\alpha(t)\big\rangle \notag\\&+\beta\big[\big\langle\alpha^{*^{2}}(t)\big\rangle +
\big\langle\alpha^{2}(t)\big\rangle\big] + C.\end{align}

In relation to Eqs. \eqref{dt22}, it is found to be appealing writing
\begin{align}\label{dt25}\frac{d}{dt}\alpha(t) = -\frac{\mu}{2}\alpha(t)+ \beta\alpha^{*}(t) + \eta(t),\end{align}
where $\eta(t)$ is stochastic noise force the correlation properties of which depends on specific nature of the system and environment. For instance, assuming the expectation value of Eq. \eqref{dt25} to have the same form as \eqref{dt22} leads to
\begin{align}\label{dt26}\langle \eta(t)\rangle = 0.\end{align}
It is also possible to verify in light of  Eqs. \eqref{dt22}, \eqref{dt23}, \eqref{dt24}, and \eqref{dt25} along with the fact that the noise force at time $t$ does not correlate with the cavity mode variables at earlier times that
\begin{align}\label{dt33}\langle \eta(t')\eta(t)\rangle =-B\delta(t-t'),\end{align}
\begin{align}\label{dt34}\langle \eta^{*}(t')\eta(t)\rangle = C\delta(t-t').\end{align}

It turns out to be convenient introducing new variables defined by
\begin{align}\label{dt35}\alpha_{\pm}(t) =\alpha^{*}(t)\pm\alpha(t)\end{align}
to derive time evolution of parameter $\alpha(t)$. With the aid of Eq. \eqref{dt25} and its complex conjugate, it is possible to see that
\begin{align}\label{dt36}{d\over dt}\alpha_{\pm}(t) =-{\lambda_{\mp}\over2}\alpha_{\pm}(t) + E^{*}(t)\pm
E(t),\end{align} 
where $\lambda_{\mp}=\mu\mp2\beta.$ Formal integration of Eq. \eqref{dt36} then results in
\begin{align}\label{dt39}\alpha(t+\tau) &= a_{+}(\tau)\alpha(t) +a_{-}(\tau)\alpha^{*}(t)\notag\\&+F_{-}(t+\tau)+F_{+}(t+\tau),\end{align}
where
\begin{align}\label{dt40}a_{\pm}(\tau)={1\over2}\left(e^{-{1\over2}\lambda_{-}\tau}
\pm e^{-{1\over2}\lambda_{+}\tau}\right),\end{align}
\begin{align}\label{dt42}F_{\pm}(t+\tau)&={1\over2}\int_{0}^{t}e^{-{1\over2}\lambda_{\mp}\tau} \notag\\&\times\big[E(t'+\tau)\pm E^{*}(t'+\tau)\big]dt'.\end{align}

Two-time correlation among various sets of parameters can be evaluated upon citing this approach. It is not difficult to see that
\begin{align}\big\langle\alpha^{*}(t)\alpha(t+\tau)\big\rangle& =a_{+}(\tau)\langle\alpha^{*}(t)\alpha(t)\rangle +\big\langle\alpha^{*}(t)F_{+}(t+\tau)\big\rangle\notag\\& +a_{-}(\tau)\big\langle\alpha^{*^{2}}(t)\big\rangle+\big\langle\alpha^{*}(t)F_{-}(t+\tau)\big\rangle.\end{align}
What remains to do in finding two-time second-order correlation function is to obtain the involved expectation values or correlations that can be carried out by using the correlation among system and reservoir variables \cite{arx} and the procedure provided in \cite{meth}.

\subsection{Coherent-state propagator}

In relation to mathematical intricacy required in manipulating operators, it is of common interest seeking for corresponding  $c$-number equation. One of such formulations that amounts to replacing the evolution operator with $c$-number function is dubbed as coherent-state propagator \cite{pra26451, jpamg347227}. It is shown that the rigor required to determine two-time second-order correlation function would be reduced when $c$-number usually linked to path integral formulation is utilized \cite{books, fey, jcp116507}. With this in mind, the way of obtaining two-time second-order correlation function would then be illustrated in light of coherent-state propagator. To do so, one can begin with arbitrary correlation function of the form
\begin{align}\label{5045}g(\tau) = \big\langle\hat{a}^{\dagger}(t +\tau)\hat{a}(t)\big\rangle= {\rm{Tr}}\big(\hat{\rho}(0)\hat{a}^{\dagger}(t+\tau)\hat{a}(t)\big).\end{align} 

As noted earlier, the time dependence can be transferred to the density operator as
\begin{align}\label{5047}g(\tau) = {\rm{Tr}}\big(\hat{\rho}(t)\hat{a}^{\dagger}(\tau)\hat{a}\big),\end{align}
where $\hat{\rho}(t)$ can be induced from $\hat{\rho}(0)$ as depicted in Eq. \eqref{5039}.
Upon introducing completeness relation for coherent state in Eq. \eqref{5047}, one can afterwards write in view of cyclic property of trace operation that
\begin{align}\label{5052}g(\tau) = \int\frac{d^{2}\alpha}{\pi}\alpha\big\langle\alpha|\hat{U}(t)
|\alpha_{0}\big\rangle\big\langle\alpha_{0}|\hat{U}^{\dagger}(t)\hat{a}^{\dagger}(\tau)|\alpha\big\rangle\end{align}
in case the initial state of the system is denoted by $|\alpha_{0}\rangle$ \cite{pr1302529, gerry}. 

In addition, expressing coherent state propagator as
\begin{align}K(\alpha,t|\beta,0) =\big\langle\alpha|\hat{U}(t)|\beta\big\rangle\end{align} 
and utilizing completeness relation for coherent state lead to
\begin{align}\label{5056}\big\langle\alpha_{0}|\hat{U}^{\dagger}(t)\hat{a}^{\dagger}(\tau)|\alpha\big\rangle
&=\int\frac{d^{2}\alpha_{2}}{\pi}\frac{d^{2}\alpha_{3}}{\pi}K^{*}(\alpha_{2},t|\alpha_{3},0)\notag\\&\times \big\langle\alpha_{0}|\alpha_{3}\big\rangle\big\langle\alpha_{2}|\hat{a}^{\dagger}(\tau)|\alpha\big\rangle,\end{align}
where $K^{*}(\alpha_{2},t|\alpha_{3},0)=\big\langle\alpha_{3}|\hat{U}^{\dagger}(t)|\alpha_{2}\big\rangle$. 

It is possible to show in the same way that
\begin{align}\label{5058}\big\langle\alpha_{2}|\hat{a}^{\dagger}(\tau)|\alpha\big\rangle =
{\rm{Tr}}\big(\hat{\rho}'(0)\hat{a}^{\dagger}(\tau)\big),\end{align} 
where  $\hat{\rho}'(0) = |\alpha\rangle\langle\alpha_{2}|$. One may note that implementing trace operation shifts the time dependence to new density operator as done before:
\begin{align}\label{5060}\big\langle\alpha_{2}|\hat{a}^{\dagger}(\tau)|\alpha\big\rangle =
{\rm{Tr}}\big(\hat{\rho}'(\tau)\hat{a}^{\dagger}\big),\end{align} 
where $\hat{\rho}'(\tau) =\hat{U}(\tau)|\alpha\rangle\langle\alpha_{2}|\hat{U}^{\dagger}(\tau)$. So inserting completeness relation for coherent state once again leads to
\begin{align}\label{5063}\langle\alpha_{2}|\hat{a}^{\dagger}(\tau)|\alpha\rangle&= \int{d^{2}\alpha_{4}\over \pi}\alpha^{*}_{4}\notag\\&\times K(\alpha_{4},\tau;\alpha,0)K^{*}(\alpha_{4},\tau;\alpha_{2},0).\end{align}

In relation to Eqs. \eqref{5052}, \eqref{5056},  and \eqref{5063}, one then gets
\begin{align}\label{5064}g(\tau) &= \int\frac{d^{2}\alpha}{\pi}\frac{d^{2}\alpha_{1}}{\pi}
\frac{d^{2}\alpha_{2}}{\pi}\frac{d^{2}\alpha_{3}}{\pi}{d^{2}\alpha_{4}\over\pi}\alpha\alpha^{*}_{4} \langle\alpha_{1}|\alpha_{0}\rangle\notag\\&\times\langle\alpha_{0}|\alpha_{3}\rangle
 K(\alpha_{4},\tau|\alpha,\tau) K^{*}(\alpha_{4},\tau|\alpha_{2},0)\notag\\&\times
K(\alpha,t|\alpha_{1},0)K^{*}(\alpha_{2},t|\alpha_{3},0).\end{align}
It is possible to deduce as the accompanying two-time second-order correlation function is evaluated
in view of coherent-state propagator. 

What remains to do to obtain two-time second-order correlation function is to extrapolate this derivation to the case with more parameters, adapt the coherent-state propagator for different variables, and then carry out the emerging integration  \cite{jcp116507}. In this direction, it has been shown that the coherent-state propagator related to quadratic Hamiltonian is usually expressed in exponential form, which makes the involved mathematical task simpler despite the number of integrations to be performed. As demonstration, following the procedure provided in \cite{pra465379}, the coherent-state propagator for parametric oscillator \cite{oc151384} and an ensemble of two-level atoms in a cavity  \cite{jmo551683}  have been shown to be represented by simple Gaussian function. It may then be appropriate to proclaim as this approach would be helpful to find two-time correlation function specially numerically, since integrating Gaussian function is mostly straightforward. 

\subsection{Quasi-statistical distribution function}

One of the methods that can be employed to determine correlation function, while $c$-number equation instead of operator equation is used to study quantum properties, is deploying the associated quasi-statistical distribution functions related to the density operator in anti-normal ordering \cite{pr40749, zp461, pr138b1566}. One of these functions is the Husimi $Q$-function that corresponds to the normal ordering of the density operator and applicable to calculate various order of moments \cite{hus, books}. It is so anticipated that the advantage rendered by this function can be manifested in reducing the rigor of obtaining two-time second-order correlation function. 

It is a well established fact that the $Q$-function that corresponds to time-dependent density operator \eqref{5039} can be expressed as
\begin{align}\label{5067}Q(\alpha, t)&=\frac{1}{\pi}\int\frac{d^{2}\alpha_{5}}{\pi} \frac{d^{2}\alpha_{6}}{\pi} \big\langle\alpha_{5}|\alpha_{0}\big\rangle\big\langle\alpha_{0}|\alpha_{6}\big\rangle \notag\\\times&K(\alpha,\tau|\alpha_{5},0)K^{*}(\alpha,t|\alpha_{6},0).\end{align}
It then possible to rewrite Eq. \eqref{5064} as
\begin{align}\label{5068}g(\tau)&= \int\frac{d^{2}\alpha}{\pi}d^{2}\alpha_{2}d^{2}\alpha_{4} ~\alpha\alpha^{*}_{4}\notag\\&\times Q'(\alpha_{4},\alpha^{*}_{4},\tau)Q(\alpha,\alpha^{*}_{2},t).\end{align} 
It is not difficult to note that the $Q$-functions in Eq. \eqref{5068} are the pertinent quasi-statistical distribution functions denoting the system typified in terms of different variables. To make use of this technique to obtain the two-time correlation, what one may need to do is finding the corresponding $Q$-function that can be expressed at different times and with different variables.

In the same manner, the time evolution of a quantum system can be directly derived from density operator in which the two-time correlation function can be expressed as
\begin{align}\label{5070}g(\tau)={\rm{Tr}}\big(\hat{a}^{\dagger}\hat{a}(\tau)\hat{\rho}(t)\big).\end{align} 
It is common knowledge that the density operator can be expanded in the normal order in light of a power series 
\begin{align}\label{5071}\hat{\rho}(t)=\sum_{l,m}C_{lm}(t)\hat{a}^{\dagger^{l}}\hat{a}^{m}.\end{align} 
Making use of coherent state completeness relation then leads to
\begin{align}\label{5072}g(\tau)= \int{d^{2}\alpha\over\pi}\sum_{l,m}C_{lm}(t){\rm{Tr}} \big(\hat{a}^{\dagger}\hat{a}(\tau)|\alpha\rangle\langle\alpha|\hat{a}^{\dagger^{l}}\hat{a}^{m}\big).\end{align}

In connection to the action of boson operators on coherent state \cite{gerry},
\begin{align}\langle\alpha|\hat{a}^{\dagger^{l}}=\alpha^{*^{l}}\langle\alpha|;~~~~ \langle\alpha|\hat{a}^{m}=\left(\alpha+{\partial\over\partial\alpha^{*}}\right)^{m}\langle\alpha|,\end{align}
one can verify that
\begin{align}\label{5075}g(\tau)&=\int{d^{2}\alpha\over\pi}\sum_{l,m}C_{lm}(t)\alpha^{*^{l}}
\left(\alpha+{\partial\over\partial\alpha^{*}}\right)^{m}\notag\\&\times {\rm{Tr}}\big(\hat{a}^{\dagger}\hat{a}(\tau)|\alpha\rangle\langle\alpha|\big).\end{align} 

On the other hand, on the basis of the fact that
\begin{align}\label{5076}Q(\alpha,\alpha^{*},t)={1\over\pi}\sum_{l,m}C_{lm}(t)\alpha^{*^{l}}\alpha^{m}\end{align}
when the operators are initially put in the normal order, it is possible to write
\begin{align}\label{5077}g(\tau)&=\int d^{2}\alpha~ Q\left(\alpha^{*}, \alpha
+{\partial\over\partial\alpha^{*}},t\right)\notag\\&\times{\rm{Tr}}\big(\hat{a}^{\dagger}\hat{a}(\tau)
|\alpha\rangle\langle\alpha|\big).\end{align} 
Upon noting
\begin{align}\label{5078}{\rm{Tr}}\big(\hat{a}^{\dagger}\hat{a}(\tau)|\alpha\big\rangle\big\langle\alpha|\big) =\alpha^{*}\big\langle\alpha|\hat{a}(\tau)|\alpha\big\rangle,\end{align}
\begin{align}\label{5079}\big\langle\alpha|\hat{a}(\tau)|\alpha\big\rangle= {\rm{Tr}}\big(\hat{a}(\tau)\hat{\rho}\big),\end{align} 
and introducing coherent state completeness relation once again, it is not hard to notice that
\begin{align}\label{5083}\big\langle\alpha|\hat{a}(\tau)|\alpha\big\rangle=\int d^{2}\beta~\beta Q(\beta,\beta^{*},\tau).\end{align} 

With the aid of Eqs. \eqref{5077} and \eqref{5083}, one may then find
\begin{align}\label{5084}g(\tau)&=\int d^{2}\alpha d^{2}\beta~ \alpha^{*}\beta\notag\\&\times Q\left(\alpha^{*},\alpha+{\partial\over\partial\alpha^{*}},t\right)Q(\beta^{*},\beta,\tau).\end{align}
The $Q$-functions in Eq. \eqref{5084} are essentially the same but describe the quantum system under consideration in terms of different variables. As noted earlier, this approach can be extrapolated to when there are more than two operators as in two-time second-order correlation and then carry out the resulting integrations.

\section{Conclusion}

Mathematical procedures that can be utilized to obtain two-time second-order correlation function in terms of stochastic differential equation, coherent-state propagator, and quasi-statistical distribution functions are derived. Since working with $c$-number equation is far more easier than the associated operator equation, it is expected that the derived results can be helpful in reducing the otherwise involving mathematical rigor. With this in mind, it is observed that the two-time second-order correlation can be determined once the corresponding coherent-state propagator or $Q$-function is known. On the basis that the involved functions are generally Gaussian in nature, calculating the two-time second-order correlation function is expected to be realizable despite the required many number of integrations.

\end{document}